# A kinetic Ising model study of dynamical correlations in confined fluids: Emergence of both fast and slow time scales

**Rajib Biswas and Biman Bagchi***

Solid State and Structural Chemistry Unit

Indian Institute of Science

Bangalore 560012, India

*Email: bbagchi@sscu.iisc.ernet.in

## ABSTRACT

Experiments and computer simulation studies have revealed existence of rich dynamics in the orientational relaxation of molecules in confined systems such as water in reverse micelles, cyclodextrin cavities and nano-tubes. Here we introduce a novel finite length one dimensional Ising model to investigate the propagation and the annihilation of dynamical correlations in finite systems and to understand the intriguing *shortening* of the orientational relaxation time that has been reported for small sized reverse micelles. In our finite sized model, the two spins at the two end cells are *oriented in the opposite directions*, to mimic the effects of surface that in real system fixes water orientation in the opposite directions. This produces opposite polarizations to propagate inside from the surface and to produce bulk-like condition at the centre. This model can be solved analytically for short chains. For long chains we solve the model numerically with Glauber spin flip dynamics (and also with Metropolis single-spin flip Monte Carlo algorithm). We show that model nicely reproduces many of the features observed in experiments. Due to the destructive interference among correlations that propagate from the surface to the core, one of the rotational relaxation time components decays faster than the bulk. In general, the relaxation of spins is non-exponential due to the interplay between various interactions. In the limit of strong coupling between the spins or in the limit of low temperature, the nature of relaxation of the spins undergoes a qualitative change with the emergence of a *homogeneous dynamics* where decay is predominantly exponential, again in agreement with experiments.

## I. INTRODUCTION

Dynamical correlations amongst molecules that are confined within small volumes are perturbed due to the surface and are very difficult to describe quantitatively. This is particularly so for water whose unique features are intimately related to the hydrogen bond network that gets modified by the surface interactions. In reverse micelles (RMs), where one can control the size of the system, one can systematically study the effects of confinement[1-8]. Recent experimental and computer simulation studies have studied water in variety of environments[1-11] such as water around bio-molecules[1-3], around micelles[4-6], in reverse micelles[7,8], in cyclodextrins[9,10], in Zeolites[11] and nano-tubes. Discussions have often focused on the relative effects of nano-confinement versus surface interactions. However, no purely theoretical study of such problems has been presented.

The initial experimental studies employing solvation dynamics[12,13] measurements reported a very slow component which could arise from the motion of water molecules trapped near the charged surfactant group or due to the motion of the solute probe itself. Subsequently a series of detail experimental studies of rotational dynamics of water has been carried out by Fayer and coworkers by using 2D ultrafast IR spectroscopy of O-H and O-D bonds[14-17]. These experiments revealed a wealth of information, particularly in the short timescale window. One of the salient features of the recent experimental results of Fayer and coworkers[16] is the observation of a rotational relaxation time component *faster* than the bulk. While the bulk rotational correlation time of O-D bond is 2.7ps, same near the core is found to be 1.5ps (for $w_0 = 10$, where $w_0$ is the characterization parameter for RMs, defined as water molecules per surfactant molecule). This significant acceleration of the relaxation rate in the core is yet to be explained.

Although there does not seem to exist any purely theoretical study of the problem, several computer simulation studies of water relaxation dynamics in RMs have been reported[18,19]. In one of the early pioneering studies, Faeder and Ladanyi[18] simulated and found that the rotational dynamics slow down by a factor of 2 to 3. Subsequently, Senapati and Berkowitz[19] have also reported that the translational and the rotational dynamics of water slow down in confined geometry of RMs with change in inter water hydrogen bonding. They have also observed that the reorientational relaxation of water in the solvation layer may slow down by three orders of magnitude which is in good agreement with the experimental results[7].

Several studies on dynamics of water in micelles have also been reported[4-6]. It is interesting to contrast the behavior of water in micelles and reverse micelles. In the case of reverse micelles the confinement may induce new effects which could be more important than the surface specific interactions. In a recent study of water structure and dynamics in the grooves of DNA it has been found that confinement can significantly distort the bulk behavior[20].

Recent experimental and theoretical studies on various confined systems have now given enough evidence for the presence of multiple time scales in confined water systems originating generally from the presence of two ensembles of water molecules[14-17]. One type is referred to as free or bulk water and the other is called bound or surface water whose motion is restricted. The latter is responsible for slow dynamics. Relative importance of the two relaxation modes depend on size and nature of the system.

In order to explain the interplay between these two limits of water, here we propose a one dimensional simple theoretical model. Our model is essentially a modified finite length Ising model. In this finite sized model, the two spins at the two end cells are *oriented in the opposite directions*, to mimic the effects of surface that in real system fixes water orientation in the opposite directions. Fortunately, this model can be solved analytically for short chains. For long chains we solve the model numerically with Glauber spin flip dynamics (and also with Metropolis single-spin flip Monte Carlo algorithm). Interestingly, this model can reproduce many of the features observed in experiments. Due to the destructive interference among correlations that propagate from the surface to the core, one of the rotational relaxation time components decays faster than the bulk. Another new result is that in the limit of strong coupling between the spins or in the limit of low temperature, the nature of relaxation of the spins undergoes a qualitative change with the emergence of a *homogeneous dynamics* where decay is predominantly exponential.

## II. THE MODEL

The key observation behind the model is that in confined systems, such as reverse micelles, the water molecules that are located diametrically opposite to each other are *orientated opposite to each other*. This is because the surface water molecules are strongly hydrogen bonded with the charged surface groups. The opposite correlations induced by spatially opposite surface groups propagate inside and are expected to annihilate each other at the center. Thus, we may expect a pool of water molecules at the center where water molecules are as free as or sometimes even freer than the bulk. Of course, for large sized reverse micelles, the center pool is expected to be bulk like any way. However, the correlations effect should be appreciable at small to intermediate sized micelles.

To model the above phenomenon in a simple way, we have taken a one dimensional Ising chain with two terminal spins fixed in opposite directions, one as up (i.e. '+1') and the other at the opposite side as down (i.e. '-1'). See **Fig. 1.** for a schematic illustration. We assume ferromagnetic interaction between neighboring spins. Any fixed spin will create an inhomogeneity in the system. This inhomogeneity will naturally decrease as one move away from the terminal towards the centre. At the centre the correlations imposed by the opposite surfaces can cancel each other, just as opposite polarizations are expected to do. So the central spin will behave like a free spin and it will relax faster.

We have studied this model by varying both the total number of spins and the temperature. The analytical and numerical calculations have been carried out using Glauber master equation[21] for various number of spin systems. The Glauber master equations of motion have been solved analytically only for smaller chains.

## III. THEORETICAL DEVELOPMENT

### A. Glauber equation of motion

The Glauber kinetic Ising model[21] consists of one dimensional array of spins. Each spin having two possible orientations identified by the spin variable $\sigma = \pm 1$. The energy of the system is given by the Ising Hamiltonian

$$H = -J\sum_{\langle ij \rangle}\sigma_i\sigma_j \quad \text{(The sum is all over nearest neighbors)} \qquad (1)$$

If the probability of a given spin state $(\sigma_1, \sigma_2, ..., \sigma_N, t)$ at time $t$ is $p(\sigma_1, \sigma_2, ..., \sigma_N, t)$ where $N$, is the total number of spins in the system and $w_j(\sigma_{j-1}, \sigma_j, \sigma_{j+1})$ be the probability per unit time that the $j$-th spin flips from $\sigma_j$ to $-\sigma_j$, while other spins remain fixed, then the rate of change of these probabilities with respect to time is assumed to be given by the Glauber master equation.

$$\frac{d}{dt}p(\sigma_1,\sigma_2,...,\sigma_N,t) = -\sum_j w_j(\sigma_{j-1},\sigma_j,\sigma_{j+1})p(\sigma_1,\sigma_2,.,\sigma_j,.,\sigma_N,t) + \sum_j w_j(\sigma_{j-1},-\sigma_j,\sigma_{j+1})p(\sigma_1,\sigma_2,.,-\sigma_j,.,\sigma_N,t) \tag{2}$$

where the transition probability $w_j(\sigma_{j-1},\sigma_j,\sigma_{j+1})$ is given by

$$w_j(\sigma_{j-1},\sigma_j,\sigma_{j+1}) = \frac{1}{2}\alpha\left[1-\frac{1}{2}\gamma\sigma_j\left(\sigma_{j-1}+\sigma_{j+1}\right)\right], \tag{3}$$

where $\alpha/2$ is the rate per unit time at which the spin makes transitions from either state to the opposite one.

Glauber has shown that for Ising model with detailed balance

$$\begin{aligned}\gamma &= \tanh\left(2J/k_BT\right)\\ &= \tanh\left(2J\beta\right)\end{aligned} \tag{4}$$

where $\beta = 1/k_BT$. Now one can define a new variable, the *expectation value* of the spin $\sigma_j(t)$ as $q_j(t)$. $q_j(t)$ is a stochastic variable of time,

$$\begin{aligned}q_j(t) &= \left\langle\sigma_j(t)\right\rangle\\ &= \sum_{\{\sigma\}}\sigma_j(t)p(\sigma_1,\sigma_2,...,\sigma_N,t)\end{aligned} \tag{5}$$

Here $\{\sigma\}$ refers to a sum over all the spin configurations.

By using the value of $w_j(\sigma_{j-1},\sigma_j,\sigma_{j+1})$ and $q_j(t)$ one can find the following equation of motion for the *k*-th spin

$$\frac{dq_k(t)}{dt} = -\alpha q_k(t) + \frac{\alpha\gamma}{2}\left\{q_{k-1}(t)+q_{k+1}(t)\right\} \tag{6}$$

Glauber solved this equation of motion with periodic boundary condition as well as with the central spin fixed.

Budimir and Skinner[22,23] extended Glauber model for boundary wall propagation. Bagchi and Chandra[24] solved the model to get an expression of frequency dependent dielectric function. All these studies considered only the periodic boundary condition.

This equation of motion can also be solved with two terminal spins fixed for short chains, as shown below.

### B. Solution for short chain lengths

**Case-I: For *N*=3**

Here $q_1(t)=1$ and $q_3(t)=-1$ (7)

Therefore the equation of motion becomes

$$\frac{dq_2(t)}{dt}=-\alpha q_2(t)+\frac{\alpha\gamma}{2}\{q_1(t)+q_3(t)\} \tag{8}$$

which is solved to obtain

$$q_2(t)=q_2(0)\exp(-\alpha t) \tag{9}$$

That is, the decay of the central spin follows the non-interacting limit. This is a nice and simple example of the destructive interference due to the interaction with the surface.

**Case-II: For *N*=4**

The boundary conditions remain the same.

$q_1(t)=1$ and $q_4(t)=-1$ (10)

Now the equations of motion become

$$\frac{dq_2(t)}{dt}=-\alpha q_2(t)+\frac{\alpha\gamma}{2}\{1+q_3(t)\} \tag{11}$$

$$\frac{dq_3(t)}{dt}=-\alpha q_3(t)+\frac{\alpha\gamma}{2}\{q_2(t)-1\} \tag{12}$$

Two solutions of the above coupled equations are

$$q_2(t) = \frac{\gamma}{\gamma+2} + \frac{1}{2} q_{23}^{+}(0)\exp\left(-\alpha\lambda_4^{-} t\right) + \frac{1}{2}\left[q_{23}^{-}(0) - \frac{2\gamma}{\gamma+2}\right]\exp\left(-\alpha\lambda_4^{+} t\right) \tag{13}$$

$$q_3(t) = -\frac{\gamma}{\gamma+2} + \frac{1}{2} q_{23}^{+}(0)\exp\left(-\alpha\lambda_4^{-} t\right) - \frac{1}{2}\left[q_{23}^{-}(0) - \frac{2\gamma}{\gamma+2}\right]\exp\left(-\alpha\lambda_4^{+} t\right) \tag{14}$$

where $q_{23}^{\pm}(0) = q_2(0) \pm q_3(0)$ (15)

and $\lambda_4^{\pm} = \left(1 \pm \frac{\gamma}{2}\right)$ (16)

From Eq. 13 and Eq. 14 it is clear that the decaying pattern of $q_2(t)$ and $q_3(t)$ have two time scales, one is faster and the other is slower than the non-interacting limit. At high temperature or low $J$ limit $\gamma \to 0$, thus $q_2(t)$ and $q_3(t)$ decay with only one time constant i.e. $1/\alpha$. However, the time scales differ by as much as a factor of three in the strong coupling limit ($\gamma$=1).

**Case-III: For *N*=5**

Boundary conditions remain the same as before. The equations of motion now become

$$\frac{dq_2(t)}{dt} = -\alpha q_2(t) + \frac{\alpha\gamma}{2}\left\{1 + q_3(t)\right\} \tag{17}$$

$$\frac{dq_3(t)}{dt} = -\alpha q_3(t) + \frac{\alpha\gamma}{2}\left\{q_2(t) + q_4(t)\right\} \tag{18}$$

$$\frac{dq_4(t)}{dt} = -\alpha q_4(t) + \frac{\alpha\gamma}{2}\left\{q_3(t) - 1\right\} \tag{19}$$

Three solutions of the above equations are as follows

$$q_2(t)=\frac{\gamma}{2}+\left[q_{24}^{-}(0)-\gamma\right]\exp(-\alpha t)+\left[\frac{1}{4}q_{24}^{+}(0)-\frac{1}{2\sqrt{2}}q_3(0)\right]\exp\left(-\alpha\lambda_5^{+}t\right)$$
$$+\left[\frac{1}{4}q_{24}^{+}(0)+\frac{1}{2\sqrt{2}}q_3(0)\right]\exp\left(-\alpha\lambda_5^{-}t\right) \tag{20}$$

$$q_3(t)=\left[\frac{1}{2}q_3(0)-\frac{1}{2\sqrt{2}}q_{24}^{+}(0)\right]\exp\left(-\alpha\lambda_5^{+}t\right)+\left[\frac{1}{2}q_3(0)+\frac{1}{2\sqrt{2}}q_{24}^{+}(0)\right]\exp\left(-\alpha\lambda_5^{-}t\right) \tag{21}$$

$$q_4(t)=-\frac{\gamma}{2}+\left[q_{24}^{-}(0)+\gamma\right]\exp(-\alpha t)+\left[\frac{1}{4}q_{24}^{+}(0)-\frac{1}{2\sqrt{2}}q_3(0)\right]\exp\left(-\alpha\lambda_5^{+}t\right)$$
$$+\left[\frac{1}{4}q_{24}^{+}(0)+\frac{1}{2\sqrt{2}}q_3(0)\right]\exp\left(-\alpha\lambda_5^{-}t\right) \tag{22}$$

where $q_{24}^{\pm}(0)=q_2(0)\pm q_4(0)$ (23)

and $\lambda_5^{\pm}=\left(1\pm\frac{\gamma}{\sqrt{2}}\right)$ (24)

From Eq.20-22 it is clear that $q_2(t)$ and $q_4(t)$ now decay with three time scales while the central spin for *N=5* decays with two time constants. In similar way, at high temperature or low *J* limit (where $\gamma$ vanishes), each and every spin decays with single time scale. However, at high *J* value or low temperature where $\gamma$ approaches unity the decay scenario is different.

## IV. CONTINUUM ANALYSIS

From Eq.6, we can present an interesting and new continuum analysis of the spin dynamics in the limit of large chain, particularly suitable for spins away from the boundary. The continuum equation of motion obtained from Eq.6 is given by

$$\frac{\partial q(x,t)}{\partial t}=-\alpha(1-\gamma)q(x,t)+\frac{a^2\alpha\gamma}{2}\frac{\partial^2 q(x,t)}{\partial x^2} \tag{25}$$

where $x$ denotes position along the chain and $a$ is the lattice spacing. This equation has the following interesting structure. At high temperature or low values of *J*, the decay is given by the free particle limit, with rate equal to $\alpha$. Since $\gamma = \tanh\left(\frac{2J}{k_B T}\right)$, in the opposite limit of large *J* or low temperature $\gamma \to 1$. In this limit the decay is governed by the following diffusion type equation.

$$\frac{\partial q(x,t)}{\partial t} = \frac{a^2\alpha}{2}\frac{\partial^2 q(x,t)}{\partial x^2}. \tag{26}$$

We solve Eq. 26 with the following boundary and initial conditions[25]

$q(x,t) = 1$ when $x = 0$,

$q(x,t) = -1$ when $x = L$, $L = (N-1)a$,

$q(x,0) = f(x)$ when $t = 0$ (Initial condition)

where $L$ is total length of the chain. The final solution is given by

$$\begin{aligned} q(x,t) = 1 - 2\frac{x}{L} - \frac{2}{\pi}\sum_{n=1}^{\infty}\Bigg[&\frac{\cos n\pi + 1}{n}\sin\frac{n\pi x}{L} \\ &+ \frac{2}{L}\sin\frac{n\pi x}{L}\int_0^L f(x')\sin\frac{n\pi x'}{L}dx'\Bigg]\exp\left(-\frac{a^2\alpha n^2\pi^2 t}{2L^2}\right) \end{aligned} \tag{27}$$

The time dependence of $q(x,t)$ is determined primarily by the first term $(n=1)$ of the sum. In this case, the long time is single exponential with rate equal to $a^2\,\alpha\pi^2/2L^2$.

For $L = 10a$, this predicts a decay that is an order of magnitude slower than that in the non-interacting limit. Experiments have indeed indicated that in certain limits, orientational relaxation of water in RM may become single exponential[14-17]. We find that this could be the case at low temperature where effects of ferromagnetic coupling is strong. We refer to this interesting behavior as *homogeneous dynamics*.

## V. NUMERICAL RESULTS AND DISCUSSIONS

We have solved the Glauber equation of motion (Eq.6) at temperature $T = 1.0J/k_BT$ , $\alpha = 1$ and $dt = 1\times10^{-3}$, with the terminal spins fixed in the opposite directions. As already mentioned, we have also employed Metropolis single-spin flip Monte Carlo algorithm and the results are almost identical to those from Glauber model. Because we could obtain analytical results from an equation of motion, we present here results obtained from Glauber model.

### A. Annihilation of opposite correlations

In **Fig. 2.** we have shown the behavior of equilibrium average spin values $\left(q_i\left(t_{eq}\right)\right)$ of each spin *i* for *N*=11 spins system. Since in our model both the terminal spins are fixed, the opposite polarization induced by these fixed spins will of course diminish as we move away from the terminal to the centre. There is an additional factor working here. It is expected that the opposite correlations will annihilate each other at the centre. **Fig. 2.** clearly depicts this feature. Therefore, the central spin can exhibit a behavior different from the bulk which is described here by imposing periodic boundary condition.

### B. Emergence of the fast time scale

We have also plotted the decay pattern of average spin values i.e. $q_i(t)$ for various conditions. The correlation of the present model with experiments can now be established from these results. Fayer and coworkers have reported that for the smallest RMs, the fast component of the two decay constants is faster than the bulk and the slow component is slower[16]. They have also shown that the relaxation behavior largely depends on the system size and for sufficiently large RMs, the relaxation behavior exactly matches that with bulk water. We have observed a similar behavior from our model. Calculated results are shown in **Figs. 3a.** and **3b**. In the case of small size RMs, we have found that the central spin exhibits an initial decay rate faster than the bulk with periodic boundary condition. Importantly, the system size dependence of the central spin decay *pattern* exhibits a crossover from an overall faster decay for the central spin in small chains to the bulk –like behavior of the same for longer chains, as shown in **Fig. 3a**. For small chains (like $N = 5$), we find that the central spin indeed has two decay constants, one is faster and the other is slower, than the bulk, as shown in **Fig. 3b**.

**Fig. 3a.** shows that the decay pattern converges to the bulk-behavior for all chains of length greater than 15 in this finite chain model. It also shows that decay of spins in the non-interacting limit (given by the rate $\alpha$) is considerably faster than that in the interacting limit. This is the reason for the faster decay of the central spin for small chains.

### C. Emergence of the slow time scale

In our model the two terminal fixed spins mimic the restriction due to surface interactions. As a result of inter-spin interactions, the spins which are close to the fixed ones also relax slowly. This behavior is seen clearly in **Figs. 4.** and **5.** under various conditions. **Fig. 4.** further shows how closely the decay behavior of the central spin could follow that of the non-interacting (NINT) limit. As already mentioned, while this explains the origin of a relaxation component faster than the bulk, the contribution of this fast relaxation component could be small in amplitude as it is confined only near the core and could be significant only for small sized systems only. Thus the effect of confinement is more dramatic in **Fig. 5**.

### D. Effect of temperature

From previous discussions it should be clear that in the limit of very high temperature or in the limit of very low *J* value, orientational relaxation approaches the non-interacting limit. **Figs. 6a.** and 6**b.** show the variation of relaxation behavior of the central spin of $N$=5 system with temperature. It is clear from the **Fig. 6b.** that above certain temperature the bimodal decaying pattern of spin indeed becomes single exponential with the decay rate identical to that of non-interacting limit.

## VI. CONCLUSION

Let us first summarize the main results of the paper. We have introduced a new variant of the kinetic Ising model in order to model the effects of nano-confinement on the orientational dynamics of a liquid. The model assumes that the two spins at the two ends of the one dimensional chain remain fixed in the opposite directions. In the results presented here we have assumed that the spins obey single flip Glauber dynamics. We have also used Metropolis single spin flip Monte Carlo algorithm to study the dynamics in a finite system and the results obtained for the schemes are similar to the ones for Glauber dynamics. The latter, because of the existence of the equation of motion, is more amenable to theoretical analysis and largely followed here. We find the emergence of multiple time scales in the orientational dynamics as the chain length increases.

We find two results of particular interest in the context of recent experiments on reverse micelles (RMs). First, for small to intermediate sized chains, the orientational dynamics of spins at the center acquire a decay component *which is faster than the bulk*. This rapid decay component is a result of the cancellation of the polarization caging propagating from two ends of the chain. This result may provide a simple explanation of a similar acceleration of orientational decay observed for water molecules in the central pool of RM by Fayer and coworkers where the decay time constant was found to shorten from 2.7 ps for bulk to 1.5 ps in an intermediate sized RM.

The second interesting result is the emergence of a single homogeneous decay behavior in the strong coupling or low temperature limit. In this limit we could perform a continuum model analysis. In the limit of low temperature or strong coupling, dynamics of each spin is shown to follow a diffusion equation. This can be solved and a single homogeneous decay behavior is obtained. This is again in agreement with recent experiments. The time constant of the decay was found to depend on the length of the chain in an interesting fashion.

The variant of the kinetic Ising model studied here seems to have not been investigated earlier. All the earlier models (Glauber, Skinner, Bagchi and Chandra) employed periodic boundary condition. Fortunately, we could solve the problem in different limiting situations analytically even in the absence of the periodic boundary condition.

It is straight forward to generalize this work to allow for orientation of the spins. Such a relaxation will reduce the effects of confinement.

**ACKNOWLEDGEMENTS**

The work has been supported by grants from DST and CSIR, India. B. B. acknowledges support from J. C. Bose fellowship (DST) and R. B. acknowledges CSIR for research fellowship.

**List of Figures**

**FIG. 1.**

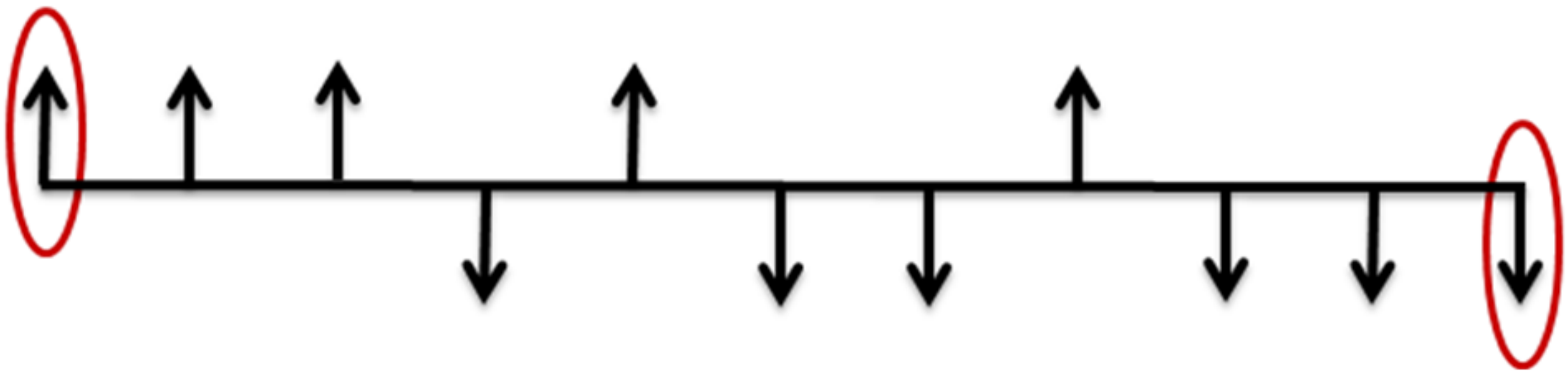

**FIG. 2.**

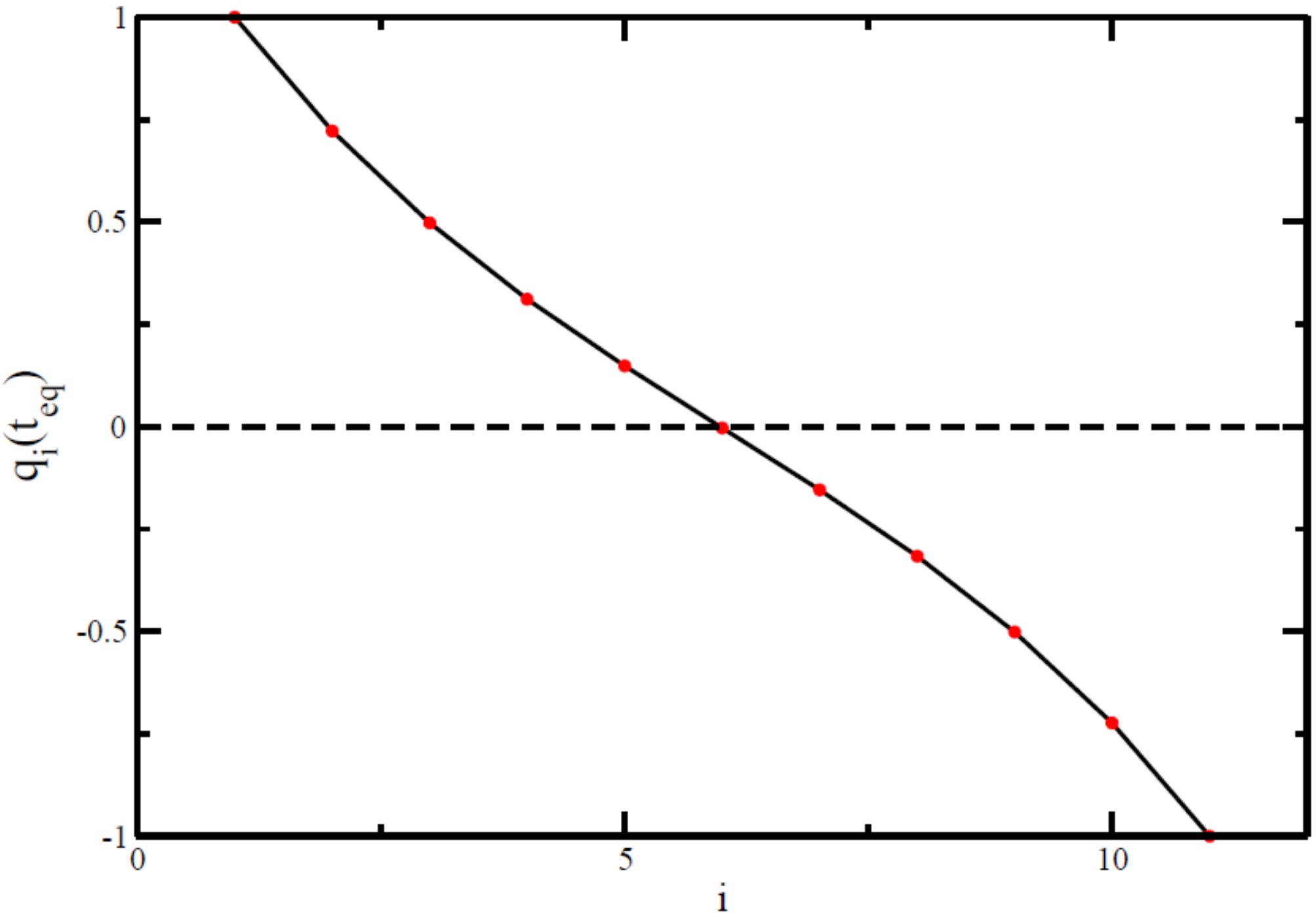

**FIG. 3a.**

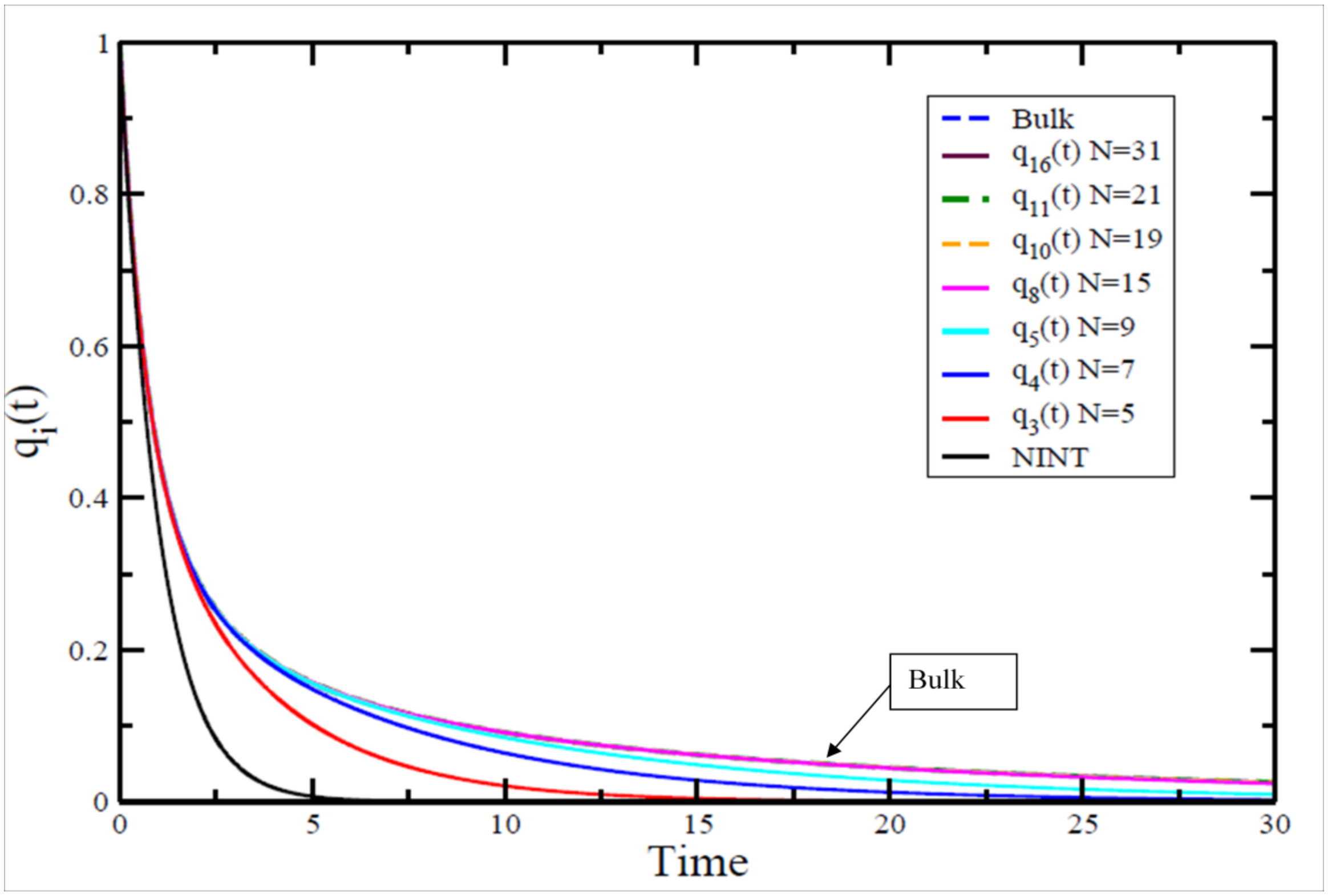

1
0.8
0.6
0.4
0.2
0
0
5
10
15
20
25
30
Time
$q_i(t)$
Bulk
$q_{16}(t)$ N=31
$q_{11}(t)$ N=21
$q_{10}(t)$ N=19
$q_8(t)$ N=15
$q_5(t)$ N=9
$q_4(t)$ N=7
$q_3(t)$ N=5
NINT
Bulk

**FIG. 3b.**

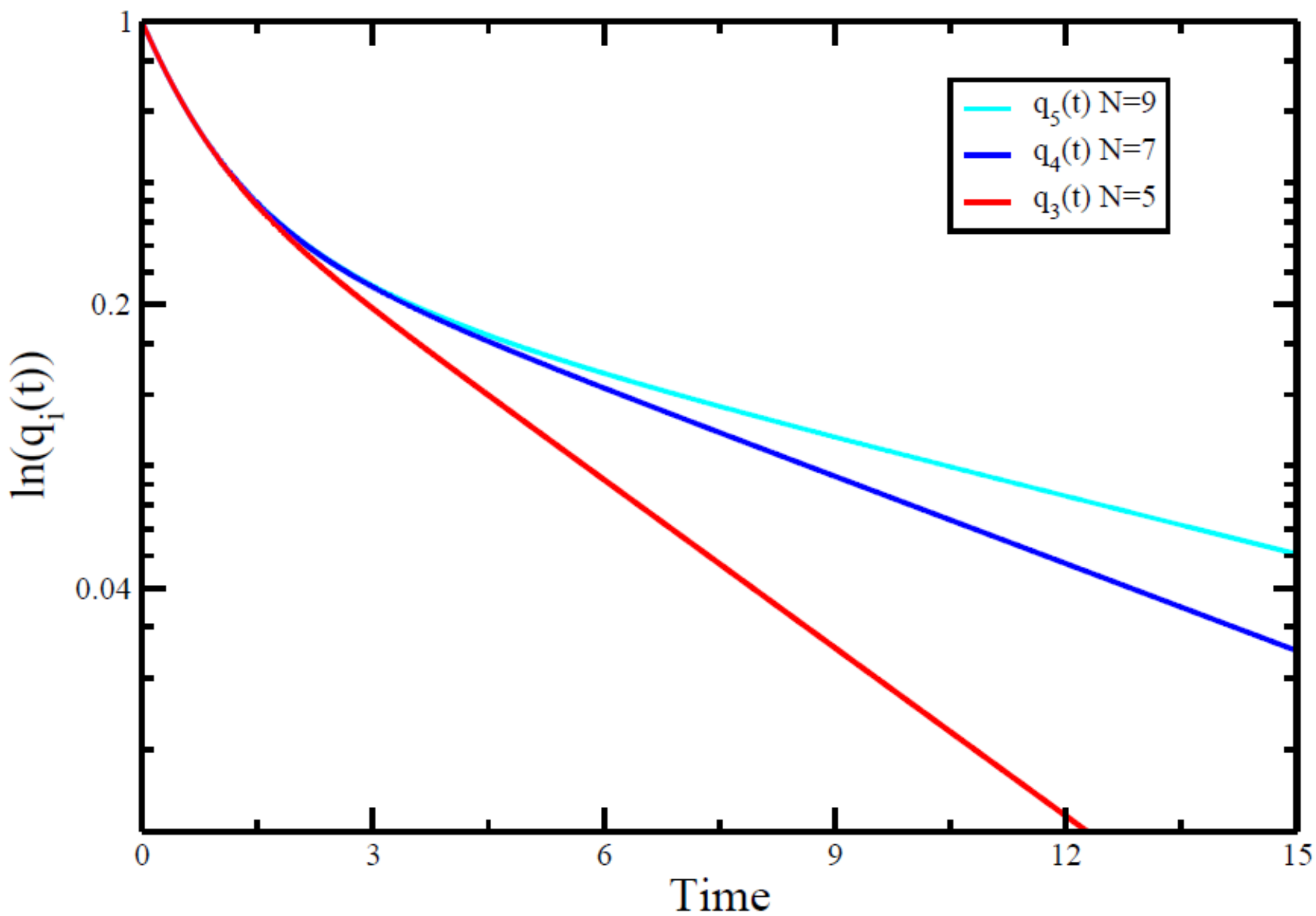

**FIG. 4.**

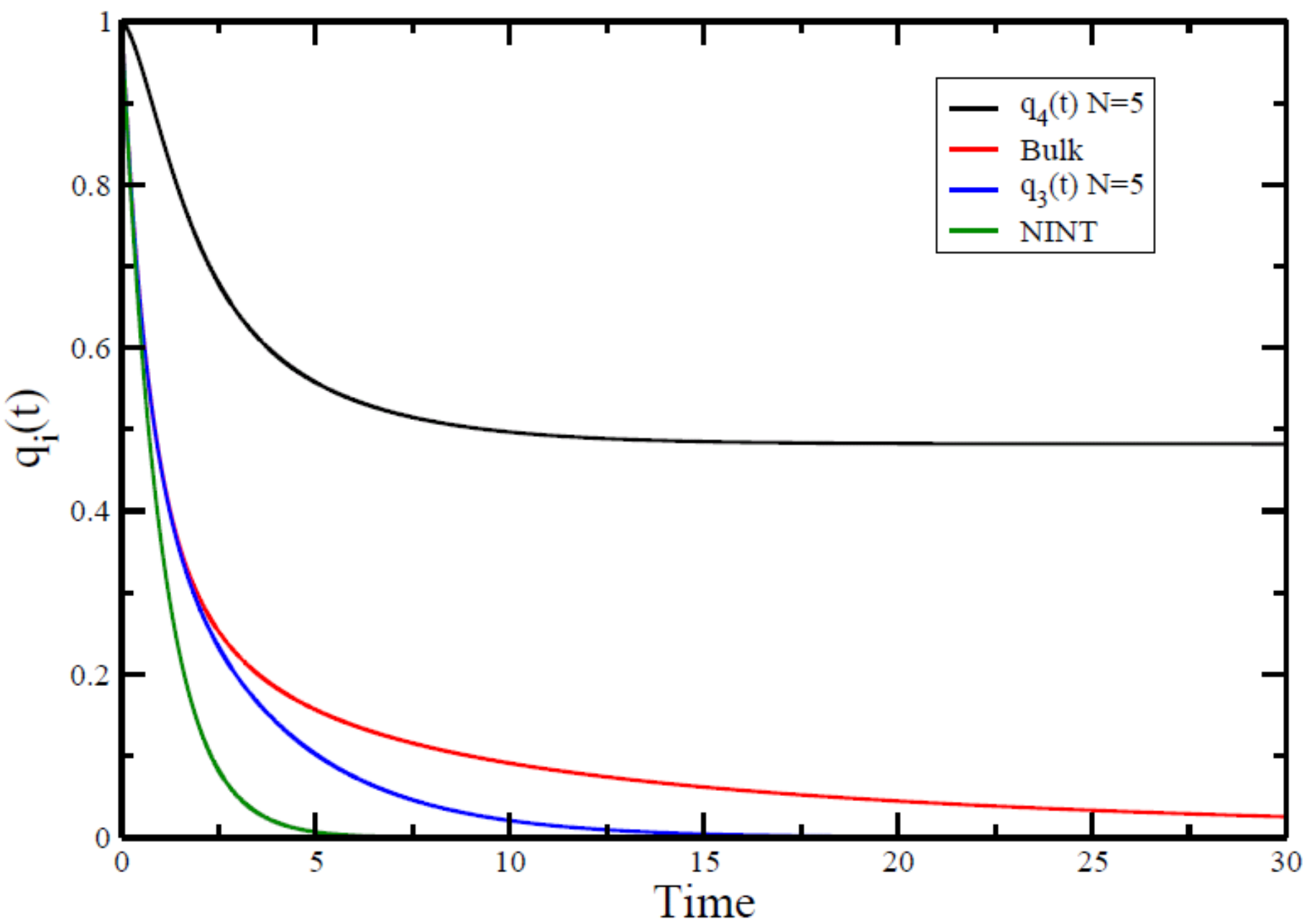

**FIG. 5.**

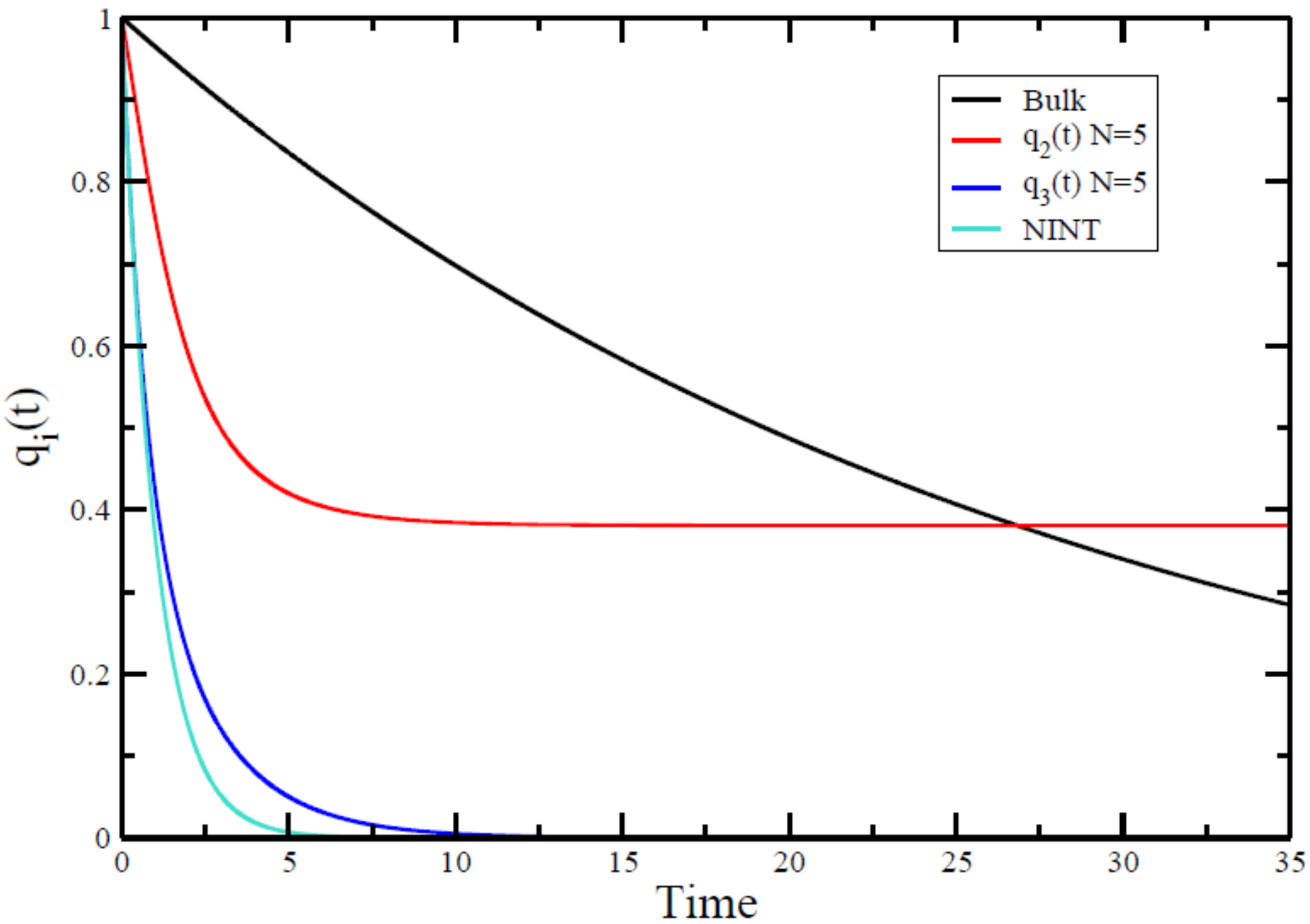

**FIG. 6a.**

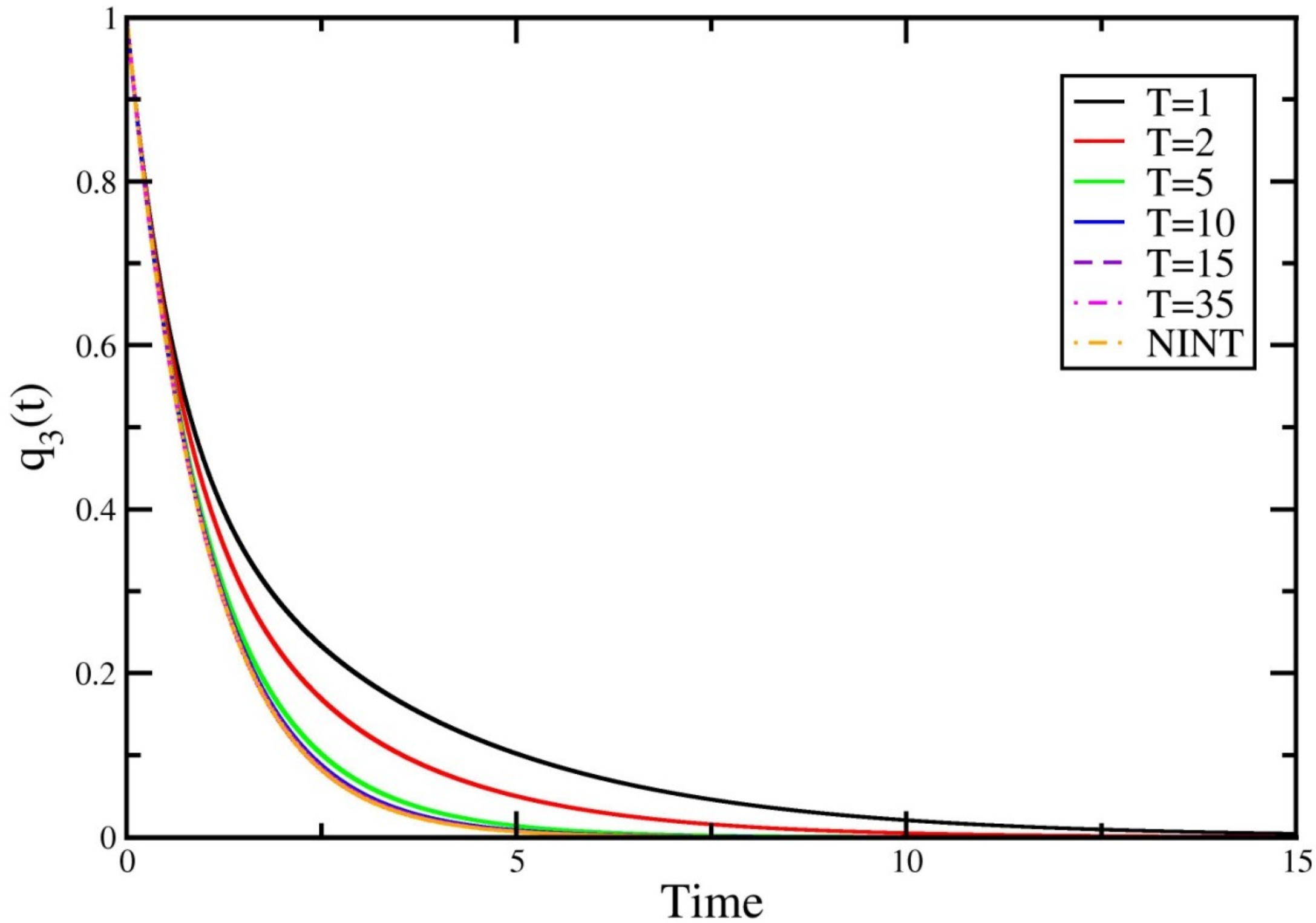

**FIG. 6b.**

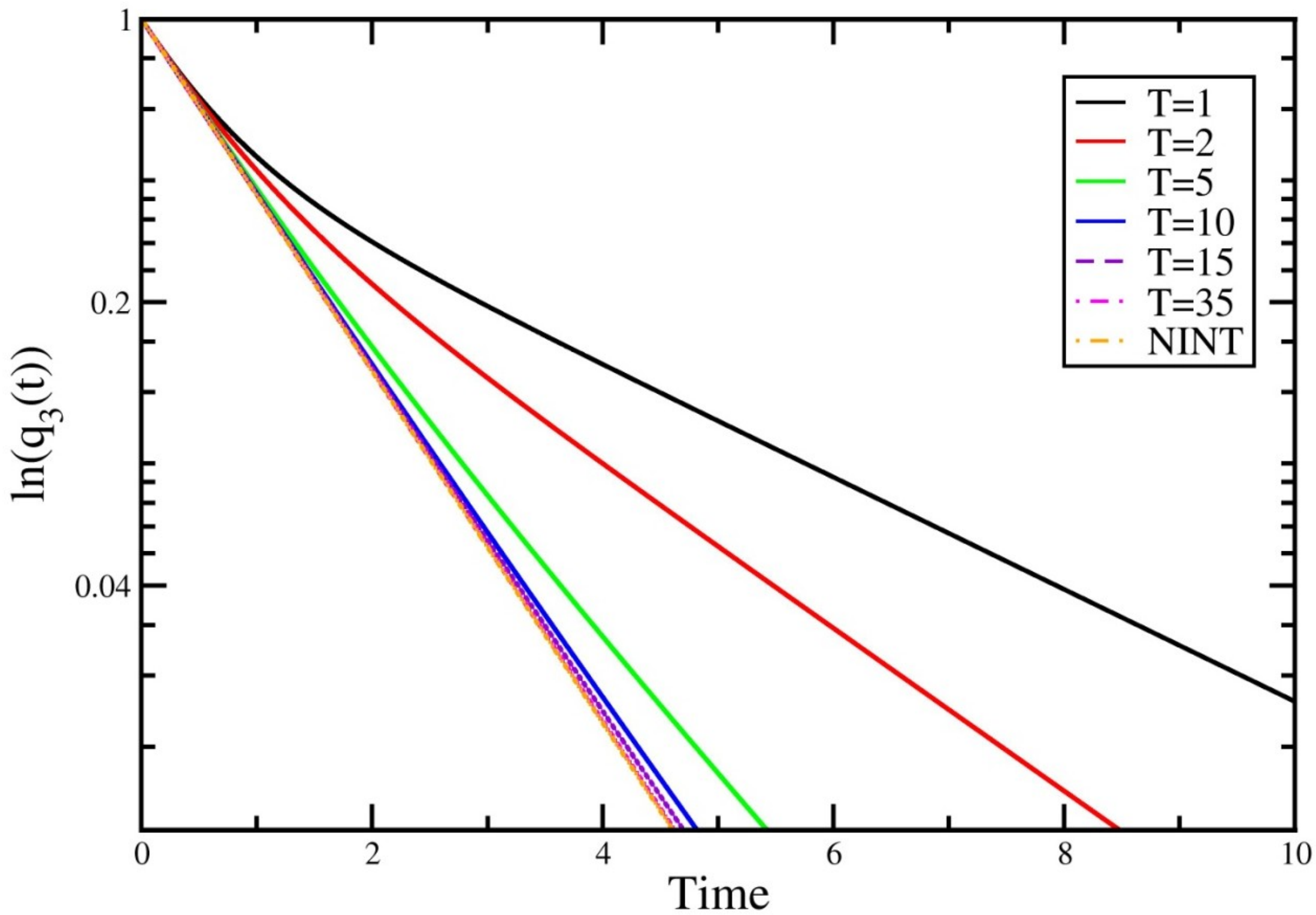

## List of figure captions

**FIG. 1. A schematic illustration of our one dimensional Ising chain model with the spins at the boundary having fixed orientation in the opposite directions.**

**FIG. 2. The calculated equilibrium average spin values for various spins for a *N*=11 chain system along the chain. Note the zero average value for the central spin (N=6).**

**FIG. 3a. A comparative plot of the calculated time dependent average spins $q_i(t)$, for various spins with following notation. The central spin (the middle one) for different chain lengths (*N* = 5 (red, solid line), 7 (blue, solid line), 9 (cyan, solid line), 15 (magenta, solid line), 19 (orange, dashed line), 21 (green, dashed line), 31 (maroon, solid line)), bulk (blue, dashed line) and non interacting case (black, solid line) with temperature $T = 1.0J/k_BT$, $\alpha = 1$ and $dt = 1\times10^{-3}$.**

**FIG. 3b. The plot of $\ln q_i(t)$ vs. time of central spins for different chain lengths; *N* = 5 (red), 7 (blue), 9 (cyan). The other parameters are the same as in Fig. 3(a).**

**FIG. 4. The plot of the calculated time dependent average spins $q_i(t)$, for various spins: a) *i* = 3 (blue), b) *i* = 4 (black) of *N*=5 spin system, c) bulk (red) and d) non interacting (green). The other parameters are the same as in Fig. 3(a).**

**FIG. 5. The plot of the calculated time dependent average spins $q_i(t)$, averaged over initial condition in various cases: a) *i* = 2 (red), b) *i* = 3 (blue) of *N*=5 spin system, c) bulk (black) and d) non interacting (cyan). The other parameters are the same as in Fig. 3(a).**

**FIG. 6a. The plot of the calculated time dependent average spins $q_3(t)$ i.e. the central spin of $N = 5$ system at various temperatures; T=1 (black solid line), T=2 (red solid line), T=5 (green solid line), T=10 (blue solid line), T=15 (violet dashed line), T=35 (magenta dashed line) and non-interacting case (orange dashed line). The other parameters are the same as in Fig. 3(a).**

**FIG. 6b. The plot of $\ln q_3(t)$ vs. time of the central spin of $N = 5$ system at various temperatures; T=1 (black solid line), T=2 (red solid line), T=5 (green solid line), T=10 (blue solid line), T=15 (violet dashed line), T=35 (magenta dashed line) and non-interacting case (orange dashed line). The other parameters are the same as in Fig. 3(a).**